\documentclass[prb,twocolumn,amsmath,amssymb,aps
]{revtex4-2}
\usepackage[normalem]{ulem}
\usepackage{lipsum}
\usepackage{graphicx}
\usepackage{dcolumn}
\usepackage{bm}
\usepackage{bbm}
\usepackage[
  colorlinks=true,
  linkcolor=magenta,   
  citecolor=blue,      
  urlcolor=teal        
]{hyperref}
\makeatletter
\newcommand*{\rom}[1]{\expandafter\@slowromancap\romannumeral #1@}
\DeclareMathOperator{\sech}{sech}
\usepackage {xcolor}
\usepackage{cancel}

\def\bef{\begin{framed}}
	\def\eef{\end{framed}}
\def\be{\begin{equation}}
	\def\ee{\end{equation}}
\def\ber{\begin{eqnarray}}
	\def\eer{\end{eqnarray}}

\def\nn{\nonumber}

\begin{document}
	
	\title{Two-Dimensional Twisted Ferromagnetic Domain Wall as a Spin-Wave Diffraction Grating}
	\date{\today}
	
	\author{Ehsan Faridi$^{1,2}$}  
	\email{faridie@savannahstate.edu}
	\author{Se Kwon Kim$^3$}
	\author{Giovanni Vignale$^{1,4}$}
	\email{vignaleg@missouri.edu}

	\affiliation{$^1$Department of Physics and Astronomy, University of Missouri, Columbia, Missouri 65211, USA\\$^2$ Department of Science and Engineering Technology, Savannah State University, Savannah, GA 31404\\$^3$Department of Physics, Korea Advanced Institute of Science and Technology, Daejeon 34141, Republic of Korea\\$^4$ The Institute for Functional Intelligent Materials (I-FIM), National University of Singapore, 4 Science Drive 2, Singapore 117544}

	\begin{abstract}
  We present a theoretical study of spin-wave scattering by a twisted domain wall (DW) in a two-dimensional ferromagnet with easy-axis anisotropy. While the twisted DW generates an effective gauge field for spin waves, leading to a deflection of their trajectories, our main focus is on a distinct effect that arises when a hard-axis anisotropy is present in addition to the easy-axis anisotropy. In this case, the translational symmetry of the spin-wave Hamiltonian along the DW is broken, resulting in a periodic modulation of the Hamiltonian. This periodicity leads to the formation of multiple diffracted spin wave modes on both sides of the DW, engendering a DW-induced magnonic diffraction pattern. The interplay between the emergent gauge field and the anisotropy-induced periodicity reveals rich spin-wave dynamics and suggests potential applications for manipulating magnon flow in two-dimensional magnetic textures.
		 
	\end{abstract}
	
	\maketitle
	
 \textit{Introduction}.|When a monochromatic beam of light encounters an obstacle, it is generally scattered, with the angular distribution of the scattered intensity determined by the geometry and material properties of the obstacle. In certain special cases, however, when the obstacle has structure on the scale of the wavelength---such as a periodic grating---the light is scattered into multiple distinct wave vectors which are called diffraction orders. The interference of these components produces a diffraction pattern, a characteristic spatial distribution of intensity whose positions and relative weights are determined by the underlying periodic structure and by the boundary conditions at the interfaces \cite{loewen2018diffraction, bonod2016diffraction}.
 
 Similarly, a spin wave---a collective excitation of magnetic order in magnetic materials---can produce diffraction patterns that arise from the coherent interference of spin-wave modes scattered by the underlying magnetic textures  \cite{mansfeld2012spin, temdie2024probing, vlaminck2023spin}. For instance, it has been numerically shown that the change in the phase acquired by the spin waves after passing through DW can be used to produce interference effects by splitting the waves on different branches of a ring \cite{hertel2004domain}. Spin-wave interference has also been demonstrated in micromagnetic simulations using geometries that mimic Young’s double-slit experiment \cite{choi2006spin}. Such interference effects have been proposed as the basis for nanoscale spectrum analyzers \cite{papp2017nanoscale}.  More recently, experiments have reported interference patterns originating from the superposition of spin waves with non-uniform amplitude profiles, where the pattern can be tuned by adjusting the composition of the magnetic system \cite{girardi2024three}. 
These examples show that spin-wave diffraction and interference are promising for magnonic devices; however, in most cases, the diffraction pattern is fixed by the geometry or material parameters which cannot be easily changed \cite{liu2022control, liu2023spin, jiang2025realizing}.

	
In this letter, we show that when spin waves scatter from a twisted DW in the presence of in-plane anisotropy form a diffraction pattern that can be controlled by changing the twist of the DW through the boundary conditions.

\textit{Twisted DW}.\,---\, Our model is a quasi-two-dimensional biaxial ferromagnet with easy-axis anisotropy normal to the plane and hard-axis anisotropy lying in the plane. Well below the Curie temperature, the magnet’s low-energy dynamics are captured by the direction of the local spin density $\textbf{m}$. The Lagrangian density in a 2D system is given by \cite{chandra1990quantum}: 

\begin{equation}
	\mathcal{L}=\mathcal{J}~\textbf{a(m)} \cdot\dot{\textbf{m}} - \mathcal{U(\textbf{m})}\, .\label{Lagrangian1}
\end{equation}

Here, the state of the ferromagnet is described by the order parameter $ \mathbf{m}(\mathbf{x}, t)=(\sin\theta\cos\phi,\sin\theta\sin\phi,\cos\theta)$, parameterized by two scalar fields: $\theta = \theta(\textbf{x},t)$ and $ \phi=\phi(\textbf{x},t)$. In Eq. (\ref{Lagrangian1}), $\mathcal{J}$ 
is the scalar density of angular momentum and  $\textbf{a}(\textbf{m})=\frac{(1-\cos\theta)}{\sin\theta}\hat{\boldsymbol{\phi}}$
is a vector potential responsible for spin dynamics. Here $\hat{\boldsymbol{\phi}}$ denotes the unit vector in the azimuthal direction in spin space \cite{kim2023mechanics,altland2010condensed}.  The second term  gives the potential energy density functional, 
\be\mathcal{U}(\textbf{m})=A(\nabla \textbf{m})^2 - K_zm_z^2 + K_x m_x^2\,.
\ee
Here, the first term represents the exchange energy density with exchange stiffness $A$; 
the second term corresponds to the easy-axis anisotropy energy density with constant $K_z$, 
favoring spin alignment along the $z$-axis; and the last term describes the hard-axis anisotropy with constant $K_x$, 
  penalizing spin alignment along the $x$-axis.  

After plugging $\textbf{m}$ into Eq. (\ref{Lagrangian1}),  the    Lagrangian density can be expressed in terms of the  two fields $\theta$ and $\phi$:

\ber
	\mathcal{L}&=& \mathcal{J}\dot{\phi}(1-\cos\theta)-\frac{A}{2} (\nabla\theta)^2 -\frac{A}{2}  \sin^2\theta(\nabla\phi)^2 \nn\\ &+& \frac{K_z}{2}\cos^2\theta - \frac{K_x}{2}\sin^2\theta\cos^2\phi  \, .\label{Lagrangian2}
\eer

\begin{figure*}
	\centering
\includegraphics[width=170mm ]{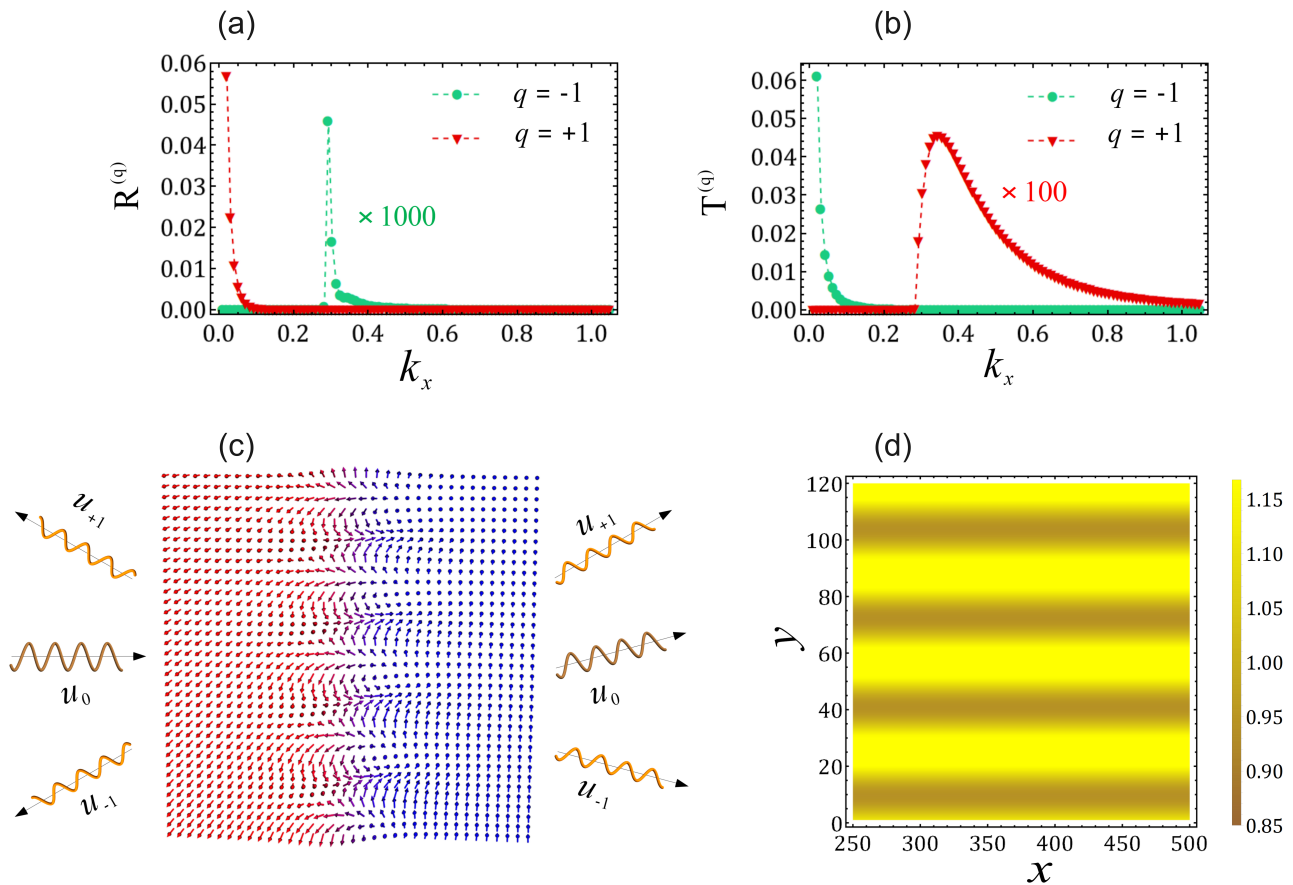}\caption{    (a)  Reflection coefficients, $|\frac{u_q(x\to-\infty)}{u_0(x\to-\infty)}|^2$ of diffracted wave as a function of $k_x$. (b) Transmission coefficients, $|\frac{u_q(x\to+\infty)}{u_0(x\to+\infty)}|^2$ of diffracted wave as a function of $k_x$. The  parameters are $\mathcal{K}_z=0.01$, $k_0=0.1$, $\mathcal{K}_x=0.001$ and $k_y=0$. (c) Schematic illustration of the interaction between a spin wave and a twisted DW. Note the change in the transverse wave vector of the transmitted spin waves caused by the twisting of the DW. (d) Density plot of $|\psi(x,y)|^2$  in the homogeneous region on the right side of the DW, as a function of $x$ and $y$ for  $\mathcal{K}=0.001$,  $\mathcal{K}_z=0.01$, $k_0=0.1$, $k_x=0.05$ and $k_y=0$. }
\label{Fig5_3}
\end{figure*}

\begin{widetext}
    
Using the Euler-Lagrange equation, 	$\frac{\partial \mathcal{L}}{\partial\varphi}-\partial_{\mu} ( \frac{\partial \mathcal{L}}{\partial(\partial_{\mu}\varphi)}  )=0$, where $\varphi=\theta,\phi$ and $\mu\in \{t, x , y\}$,  we obtain the equations governing the magnetization dynamics: 
 
 \begin{subequations}
 	\begin{align}
 		\mathcal{J}\,\dot{\phi}\,\sin\theta
 		+ A\nabla^{2}\theta
 		- \tfrac{1}{2}\sin(2\theta)\!\left[A(\nabla\phi)^{2}
        +K_{z}+K_{x}\cos^{2}\phi\right]
 		= 0, \label{eq:3a}\\
 		-\mathcal{J}\,\dot{\theta}\,\sin\theta
 		+ A\,\nabla\!\cdot\!\big(\sin^{2}\theta\,\nabla\phi\big)
 		+ \tfrac{1}{2}K_{x}\sin^{2}\theta\,\sin(2\phi)
 		= 0. \label{eq:5b}
 	\end{align}
    \label{Eq5}
 \end{subequations}
 \end{widetext}

Equation~(\ref{eq:5b}) can be recast as a continuity equation for the $z$-component of the spin density and the spin current:  
\begin{equation}
\frac{\partial \rho}{\partial t} + \nabla \cdot \mathbf{j} = -\frac{1}{2}K_x\sin^2\theta\sin 2\phi \, .\label{phiK}
\end{equation}
where $\rho = \mathcal{J} \cos\theta$ is  the local $z$-component of the spin density and $\mathbf{j} = A \sin^2\theta \nabla \phi$ is the corresponding of the spin current. The term on the right-hand side of Eq. (\ref{phiK}) arises from the in-plane anisotropy, indicating that spin angular momentum is not  conserved due to the breaking of U(1) spin-rotational symmetry of the Lagrangian about $z-$axis. A static DW that connects two discrete vacua $\mathbf{m}(x,y)=\pm\hat{\mathbf{z}}$,  can be obtained by setting $\dot{\theta}=\dot{\phi}=0$ in Eqs.~(\ref{Eq5}).
 Assuming that  $\phi$ varies only in $y-$direction, while $\theta$ varies primarily along $x$ (with slow $y-$dependence through $\phi$), then the solution for static DW is

\be
\tan\frac{\theta_0(x,y)}{2} = \exp\{\sqrt{k_0^2 +  \mathcal{K}_z+\mathcal{K}_x\cos[2\phi_0(y)]}~~x\}\, ,\label{theta_sol}
\ee

and 

\be
\phi_0(y)\approx k_0y-\dfrac{\mathcal{K}_x}{8k_0^2}\sin(2k_0y) \label{phistructureq}\, ,
\ee	

where $\mathcal{K}_{x(z)}\equiv \frac{K_{x(z)}}{A}$. In deriving Eq.~(\ref{theta_sol}) and Eq.~(\ref{phistructureq}), we assumed that  $\mathcal{K}_x\ll k_0^2\ll 1$, implying that both $\phi$ and $\theta$ vary slowly along the $y$ direction.     
As a result, the terms such as $\partial_y\theta  \partial_y\phi$ and $\partial^2_y\theta$ were neglected.	Here  $k_0\equiv \left.\frac{\partial\phi}{\partial y}\right\vert_{y=0} $ is a real number that characterizes chirality or twist of the DW texture. Variation of the DW along $y$-direction gives rise to a finite skyrmion charge density causing a deflection of magnons after passing through the DW \cite{lee2023magnon}. The DW structure of the two-dimensional spin texture, described by Eqs.~(\ref{theta_sol}) and~(\ref{phistructureq}), is one of our main results.

We analyze spin-wave excitations by introducing small time-dependent deviations from the static textured DW configuration, treating the magnetization as $\mathbf{m}(\textbf{x}, t) = \mathbf{m}_0(\textbf{x}) + \delta\mathbf{m}(\textbf{x}, t)$, where $\mathbf{m}_0(\textbf{x})$ denotes the equilibrium background given in Eq.~(\ref{theta_sol}) and Eq.~(\ref{phistructureq}). We expand the Euler--Lagrange equations, Eq.~(\ref{Eq5}), 
around the background configuration with small deviations 
$(\theta, \phi) \to (\theta + \delta\theta, \phi + \delta\phi)$. After keeping   terms linear in $\delta\theta$ and $\delta\phi$, 
we also set $\delta m_{\theta} = \delta\theta$ and $\delta m_{\phi} = \sin\theta\, \delta\phi$ as the local polar and azimuthal components of the magnetization fluctuation. 
This yields two coupled equations for $\delta m_{\theta}$ and $\delta m_{\phi}$, 
which can then be combined as $\Psi(t;x,y)= (\delta m_{\theta} - i\delta m_{\phi}, \delta m_{\theta} + i\delta m_{\phi})^T$. Considering  $\Psi(t;x,y)=e^{i\omega t}\Psi(x,y)$, we obtain: 
\begin{widetext}
\begin{align}
\frac{\mathcal{J}\omega}{A}\Psi(x,y)
&= \Big\{ \left[-\partial_x^2-\partial_y^2+\tfrac{1}{\Delta^2}\left[1-2\sech^2\left(\frac{x}{\Delta}\right)\right]\right]\sigma_z
+2ik_0\tanh\left(\frac{x}{\Delta}\right)\,\partial_y \mathbbm{1} \Big\}\Psi(x,y) \nonumber \\
&\quad +\mathcal{K}_x\Big\{-\tfrac{i}{2k_0}\tanh\left(\frac{x}{\Delta}\right)\,\partial_y \mathbbm{1}
+\tfrac{i}{2}\sin(2k_0y)\tanh\left(\frac{x}{\Delta}\right)(\mathbbm{1}-\sigma_x) \nonumber \\
&\qquad\quad +\cos(2k_0y)\big(\tfrac{i}{2k_0}\tanh\left(\frac{x}{\Delta}\right)\,\partial_y \mathbbm{1}
+\tfrac{1}{2}(i\sigma_y+\sigma_z)-2\sech^2\left(\frac{x}{\Delta}\right)\sigma_z\big)\Big\}\Psi(x,y),
\label{swhamiltonian}
\end{align}
\end{widetext}

 where $\sigma_i$ denote the Pauli matrices, $\mathbbm{1}$ is the identity matrix,  and $\Delta=1/\sqrt{\mathcal{K}_z+k_0^2}$ is the  characteristic length of the
chiral DW.  We define the operator appearing on the right-hand side of Eq.~(\ref{swhamiltonian}) as the spin-wave Hamiltonian $\mathcal{H}_\mathrm{sw}$, acting on the two-component spinor $\Psi(x,y)$ and satisfying $\mathcal{H}_\mathrm{sw}~\Psi(x,y)=\frac{\mathcal{J}\omega}{A}~\Psi(x,y)$, where the eigenvalues determine the allowed spin-wave frequencies $\omega$ on top of the twisted DW. The presence of non-diagonal matrices $\sigma_x$ and $\sigma_y$  in Eq.~(\ref{swhamiltonian}) indicates that the in-plane anisotropy couples $\delta m_{\theta}$ and $\delta m_{\phi}$, leading to the formation of elliptical spin waves. 
  Note that the spin-wave Hamiltonian is periodic in $y$ direction with the periodicity of $\pi/k_0$. Therefore, the wave functions can be expressed as, 
 \be
 \Psi_{k_y} (x,y)=\sum_q u_{k_y,q}(x)e^{i(k_y+2k_0q)y} \, ,\label{wavefunction2}
 \ee

  where  $k_y+2qk_0$ is the effective  transverse momentum. Substituting Eq. (\ref{wavefunction2}) in the Eq. (\ref{swhamiltonian}) we get:


\ber
\frac{\mathcal{J}\omega}{A}\,&u_q(x)& \;=\; H_q(x)\,u_q(x)\nn\\ &+& \mathcal{K}\,\Big( \mathcal{M}_+(q,x)\,u_{q+1}(x) \;+\; \mathcal{M}_-(q,x)\,u_{q-1}(x) \Big).\nn\\ \label{rec}
\eer

  This equation represents a set of  coupled equations for the (infinitely many) amplitudes  $u_{k_y,q}(x)$. Here $\mathcal{M}_{+}$ and $ \mathcal{M}_{-}$ denote the coupling potentials that connect the mode $q$ to its neighboring modes $q+1$ and $q-1$, respectively. In the absence of in-plane anisotrpy i.e. $\mathcal{K}_x=0$, only the $q=0$ mode contributes in the scattering process. We interpret the equation as the Schrödinger equation, $\frac{\mathcal{J}\omega}{A}\,u_0(x)=H_0(x)\,u_0(x)$, where   $u_0(x)$ describes the unperturbed dynamics of the $0-$th  Fourier component and $H_0(x)$ is the unperturbed Hamiltonian contains the  kinetic energy $-\partial^2$ and potential energy given by

\begin{align}
   V(x)=   k_y^2 +  \frac{1}{\Delta^2} \left[1-2\sech^2\left(\frac{x}{\Delta}\right)\right] + 2k_0 k_y\tanh\left(\frac{x}{\Delta}\right)\,.   \label{Schrödinger1D}
\end{align}

\bigskip

 The potential energy $V(x)$ corresponds to the Rosen--Morse potential, for which the Schrödinger equation is exactly solvable \cite{gadella2017hyperbolic, faridi2024prresearch043047}. The first component of the spinor $u_0(x)$ describes a propagating wave, whereas the second component represents an evanescent mode that is localized near the DW and decays exponentially with distance \cite{faridi2022atomic, kim2023interaction, faridi2025thermal}.


 The nontrivial twisting of the magnetic texture generates an emergent magnetic field, which  influences the magnon dynamics through an effective Lorentz force given by $\mathbf{B}  
= -\mathcal{J}\,\mathbf{m}\cdot (\partial_x \mathbf{m} \times \partial_y \mathbf{m})\hat{z}$  \cite{tatara2019effective}.

Let us first consider the case $\mathcal{K}_x = 0$, which corresponds to a uniform effective magnetic field along the $z$ direction. For a magnon incident on the DW, the transverse momentum changes due to the emergent Lorentz force. This change can be understood by distinguishing between the canonical momentum $\mathcal{J}k_y$ and the kinematic momentum $p_y=\mathcal{J}k_y-\mathcal{A}_y$, where $\mathcal{A}_y=-\mathcal{J}\cos\theta\partial_y\phi_0$ is the effective vector potential associated with the texture. In the absence of any $y$-dependent modulation in $\mathcal{H}_{sw}$, i.e. $\mathcal{K}_x=0$, translational symmetry of $\mathcal{H}_{sw}$ along $y$ ensures that the canonical momentum $\mathcal{J}k_y$ is conserved. At the same time, the twisted DW texture generates a nonzero vector potential, $\mathcal{A}_y(x)$  which approaches different asymptotic values on the two sides of the DW; consequently, the asymptotic values of the kinematic momentum differ across the wall, with $p_y = \mathcal{J}(k_y - k_0)$ for $x \to -\infty$ and $p_y = \mathcal{J}(k_y + k_0)$ for $x \to +\infty$, leading to a deflection of the magnon's trajectory.


To first order in $\mathcal{K}_x$, the emergent magnetic flux is localized along the $x$ direction near the magnetic texture, while it exhibits a weak oscillatory modulation along the $y$ direction. 
  The change in the magnon's transverse momentum after passing through the DW is given by
\be
\Delta p_y = 
\int (\mathbf{v} \times \mathbf{B}) \cdot \hat{\mathbf{y}}\, dt
= 2\,\mathcal{J}k_0 
- \frac{\mathcal{J}\,\mathcal{K}_x}{2k_0}\cos(2k_0 y)\, ,\label{changesmomentum}
\ee
where $\mathbf{v} = d\mathbf{r}/dt$ is the magnon velocity \cite{kim2017magnetic}.
{When the in-plane anisotropy is present, the periodic modulation of the texture along $y$ breaks translational invariance of $\mathcal{H}_{sw}$, allowing for  an additional $y$-dependent
 contribution to the kinematic momentum, as reflected in the oscillatory term in Eq.~(\ref{changesmomentum}).}
Each magnon that crosses the wall at a given position $y$ experiences a slightly different transverse impulse; however, its average over one full period in $y$, $
\langle \Delta p_y \rangle_y = 2\mathcal{J}k_0\, ,$ remains constant.  
{From a wave-mechanical point of view, the periodic modulation along the $y$ direction acts as an effective diffraction grating for spin waves.} According to Eq.~ (\ref{rec}),  this modulation couples neighboring Fourier harmonics $q \rightarrow q \pm 1$. For weak modulation, $\mathcal{K}_x\ll k_0^2 \ll 1$, the dominant scattering out of the central channel $q=0$ occurs into the first sidebands $q = \pm 1$. To leading order, all other modes can be neglected, and the problem reduces to a closed three--mode system. Accordingly, we restrict the equation to $q \in \{-1, 0, +1\}$.

 Far from the DW, each diffraction order $q$ is characterized by an asymptotic kinematic momentum. On the left side of the wall ($x \to -\infty$), the transverse kinematic momentum is given by
$
p_y = \mathcal{J}\bigl(k_y + 2qk_0 - k_0\bigr)$,  while on the right side of the wall ($x \to +\infty$) it becomes
$
p_y = \mathcal{J}\bigl(k_y + 2qk_0 + k_0\bigr)$.
 Note that the difference in the kinematic momentum of the diffracted waves on the two sides of the DW, $2\,\mathcal{J}k_0$, is in agreement with $\langle \Delta p_y\rangle _y$ shown under Eq.~(\ref{changesmomentum}).

The first-order amplitudes  $u_{k_y,-1}(x)$ and  $u_{k_y,1}(x)$ obtained using the Born approximation are explicitly given by

\be\label{DiffractionAmplitude2}
u_{k_y,\pm1}(x)=\int\mathcal{G}_{\pm 1}(x,x') \mathcal{M}_{\mp}(0,x') u_{k_y,0}(x')dx' \, .   
\ee

where the Green's function
$\mathcal{G}_{ q }(x,x')=\left[\frac{\mathcal{J}\omega}{A}- H_q\right]^{-1}_{x,x'}$ 
(see Appendix for additional discussion).
  
  According to Fig.~\ref{Fig5_3}(a) and (b), the reflection coefficient for the mode $q = +1$ and the transmission coefficient for $q = -1$ vanish when the incident wave vector is below certain values. This occurs because there is no magnon state on the left side of the DW for the $q = +1$ mode, and no magnon state on the right side for $q = -1$, with the same wave vector and energy as the incoming wave with $q = 0$. Additionally, the scattering amplitudes of the mode $q=-1$ on the left side of the DW and that of the mode $q = +1$ on the right side diverge as the $x$ component of incoming wave vector $k_x\to 0$. The reason stems from the behavior of the Green's function for  $k_x\to 0$. In this limit, our perturbative method breaks down, and a non-perturbative solution is required to properly capture the physics. Similar results can be obtained for an oblique incidence with $k_y\neq0$. The appearance of spin-wave diffraction, with amplitudes given in Eq. (\ref{DiffractionAmplitude2}), is our main result.

 Fig.~\ref{Fig5_3}(d) represents the spatial distribution of spin-wave intensity, $|\psi(x,y)|^2$, where:

\be
\Psi(x,y)\propto u_0(x)  + u_{-1}(x) e^{-2ik_0 y}+u_{+1}(x)e^{+2ik_0y} \, .\label{sdasdrwed}
\ee

For $\mathcal{K}_x=0$,   the diffracted amplitudes are zero, resulting in a uniform distribution of spin wave intensity on the right side of the DW. However, for $\mathcal{K}_x\neq 0$, as shown in Fig.~\ref{Fig5_3}(d), the superposition of the unperturbed ($u_0$) and perturbed ($u_{\pm 1}$) waves, which have different phases along the $y$-direction, leads to destructive and constructive interference. As evident from Eq.~(\ref{sdasdrwed}), the phase difference between all diffracted and undiffracted waves is proportional to the chirality constant, $k_0$. This suggests that the diffraction pattern can be controlled by tuning the spin current injected through the boundary. Experimentally, the resulting spin-wave diffraction could be observed in low-damping magnetic materials such as YIG, for example, by using nitrogen-vacancy (NV) centers to detect the associated magnetic stray fields \cite{luthi2025long,osterholt2024detection}.

\bigskip	 
\bigskip

\textit{Discussion}\,-- In this work, we have demonstrated that a twisted DW in a two-dimensional easy-axis ferromagnet acts as a programmable diffraction element for spin waves. While the twisted texture generates an effective gauge field that deflects magnon trajectories, the key new effect arises from the presence of an additional hard-axis anisotropy. This term breaks the translational symmetry of the spin-wave Hamiltonian along the wall, imposing a periodic structure that generates multiple diffracted spin-wave modes on both sides of the DW. The diffraction pattern can be controlled through boundary conditions via the chirality constant  $k_0$. The magnonic  system described here integrates low-loss spin-wave transport and programmable spin-wave interference, offering potential applications in advanced magnonic computing devices.

	\begin{acknowledgments}
	S.K.K. was supported by Brain Pool Plus Program through the National Research Foundation of Korea funded by the Ministry of Science and ICT (NRF-2020H1D3A2A03099291).
    G.V. was supported by the Ministry
of Education, Singapore, under its Research Centre of
Excellence award to the Institute for Functional Intelligent
Materials (I-FIM, Project No. EDUNC-33-18-279-V12).
	\end{acknowledgments}

\onecolumngrid

\appendix\section{Fourier transform of the spin wave  Hamiltonian}

In this appendix, we present a more detailed derivation of Eq. (\ref{rec}) and its solution. We start from Eq. (\ref{swhamiltonian}), which contains two contributions: the first is translationally invariant in the $y$ direction and can be solved exactly, while the second is periodic in the $y$ direction. This periodicity implies that the wave function admits a Fourier representation along the $y$ direction, as given in Eq. (\ref{wavefunction2}). Substituting this form into Eq. (\ref{swhamiltonian}), the Hamiltonian reduces to the form shown in Eq. (\ref{rec}), where

\ber
H_q(x) =\left[-\frac{\partial^2}{\partial x^2}+(k_y+2k_0q)^2 + \frac{1}{\Delta^2}(1-2\sech^2[\frac{x}{\Delta}])\right]\sigma_z  
 +(2k_0 -\frac{\mathcal{K}}{2k_0} )(k_y+2k_0q)\tanh[\frac{x}{\Delta}]~\mathbbm{1}
\eer

The terms that connect mode $q$ to $q-1$ and $q+1$ are,

 \ber
     \mathcal{M}_-(q,x)= \left(\frac{1}{4} - \sech^2[\frac{x}{\Delta}]\right)~\sigma_z +\frac{1}{k_0}\left(k_y+2k_0(q-1)\right)\tanh[\frac{x}{\Delta}]~\mathbbm{1}
 -\frac{1}{4}\tanh[\frac{x}{\Delta}]~(\sigma_x-\mathbbm{1}) + \frac{1}{4} i\sigma_y   
 \label{fourirerhamiltonian}
 \eer

 and 

  \ber
     \mathcal{M}_+(q,x)= \left(\frac{1}{4} - \sech^2[\frac{x}{\Delta}]\right)~\sigma_z +\frac{1}{k_0}\left(k_y+2k_0(q+1)\right)\tanh[\frac{x}{\Delta}]~\mathbbm{1}
 +\frac{1}{4}\tanh[\frac{x}{\Delta}]~(\sigma_x-\mathbbm{1}) + \frac{1}{4} i\sigma_y   
 \label{fourirerhamiltonian}
 \eer

If the periodic modulation is weak, we keep only the modes with 
$q \in \{-1,0,+1\}$.

Define the vector

\[
\psi(x) =
\begin{pmatrix}
u_{-1}(x) \\
u_{0}(x) \\
u_{+1}(x)
\end{pmatrix}.
\]

The  eigenvalue problem then reads

\[
\left(
\begin{pmatrix}
H_{-1} & 0 & 0 \\
0 & H_0 & 0 \\
0 & 0 & H_{+1}
\end{pmatrix}
+ \mathcal{K}_x
\begin{pmatrix}
0 & \mathcal{M}_+(0) & 0 \\
\mathcal{M}_-(-1) & 0 & \mathcal{M}_+(+1) \\
0 & \mathcal{M}_-(0) & 0
\end{pmatrix}
\right)\psi(x)
= \frac{\mathcal{J}\omega}{A} \psi(x).
\]

Then the unperturbed $\frac{\mathcal{J}\omega}{A}\,u_0(x)=H_0(x)\,u_0(x)$ with $u_{-1}(x)=u_{+1}(x)=0$. To first order in $\mathcal{K}_x$, the sideband amplitudes are

\be 
u_{k_y,\pm1}(x)=\int\mathcal{G}_{\pm}(x,x') \mathcal{M}_{\mp}(0,x') u_{k_y,0}(x')dx' \, .  
\ee

where 

\be
\mathcal{G}_{ q }(x,x')=\left[\frac{\mathcal{J}\omega}{A}- H_q\right]^{-1}_{x,x'}=\left(\begin{array}{cccc}\frac{1}{2ik_1}e^{ik_1|x-x'|}&0\\	0&\frac{1}{2k_2}e^{-k_2|x-x'|}\end{array}\right)\, .\label{greens}
\ee

The integral is evaluated over the DW, which acts as the scattering region. In the above equation $k_1^2= \frac{\mathcal{J}\omega}{A}-\mathcal{K}_z-(k_y+2qk_0-k_0)^2$ and $k_2^2= \frac{\mathcal{J}\omega}{A}+\mathcal{K}_z+(k_y+2qk_0+k_0)^2$, denote the longitudinal wave vectors of the propagating and evanescent waves, respectively, obtained by imposing energy conservation for spin waves scattered by the DW. Furthermore $\mathcal{G}^{11}$ governs the behavior of the propagating wave, while $\mathcal{G}^{22}$ governs the behavior of the evanescent wave. In evaluating Eq.~(\ref{greens}), we compute the propagator as if the DW were absent. Specifically, we use the Green's function of the homogeneous right region ($\theta = \pi$) to calculate the amplitude of the transmitted diffracted wave, and that of the homogeneous left region ($ \theta = 0 $) to compute the amplitude of the reflected diffracted wave. Equation~(\ref{greens}) corresponds to the Green's function in the left region; the expression for the right region is analogous.

\nocite{*}

\twocolumngrid

\bibliography{./SWDiffreferences}

\end{document}